\documentclass[conference,compsoc]{IEEEtran}

\pdfoutput=1

\usepackage[nocompress]{cite}
\usepackage[pdftex]{graphicx}
\usepackage{amsmath}
\usepackage{url}
\usepackage{xspace}
\usepackage{enumitem}
\usepackage[dvipsnames]{xcolor}
\usepackage{listings}
\usepackage{scrextend}
\usepackage{hyperref}
\usepackage{comment}
\usepackage{fancyhdr}
\usepackage[advantage]{cryptocode}
\usepackage[utf8]{inputenc}
\usepackage[T1]{fontenc}
\usepackage[l2tabu]{nag}
\usepackage{microtype}
\usepackage{xcolor}
\usepackage{enumitem}
\usepackage{xspace}
\usepackage{algorithm,algorithmic}
\usepackage{listings}
\usepackage{todonotes}
\usepackage{comment}
\usepackage{stmaryrd}
\usepackage{mathpartir}
\usepackage{tikz}
\usepackage{jasmin}
\usepackage{booktabs}
\usepackage{multicol}
\usepackage{balance}
\usepackage[abbreviations = false]{siunitx}
\usepackage{wrapfig}
\usepackage{amssymb}

\DeclareUnicodeCharacter{00B7}{\ensuremath\cdot}
\DeclareUnicodeCharacter{03F1}{\ensuremath\rho}
\DeclareUnicodeCharacter{2192}{\ensuremath\rightarrow}
\DeclareUnicodeCharacter{21D2}{\ensuremath\Rightarrow}
\DeclareUnicodeCharacter{2200}{\ensuremath\forall}
\DeclareUnicodeCharacter{2203}{\ensuremath\exists}
\DeclareUnicodeCharacter{2208}{\ensuremath\in}
\DeclareUnicodeCharacter{202F}{~}

\usepackage{booktabs}   
\usepackage{subcaption}

\newcommand{\Jasmin}{\textsf{Jasmin}\xspace}

\newcommand{\Coq}{\textsf{Coq}\xspace}

\newcommand{\CompCert}{\textsf{CompCert}\xspace}

\newcommand{\openSSL}{\textsf{OpenSSL}\xspace}

\newcommand{\EasyCrypt}{\textsf{EasyCrypt}\xspace}

\newcommand{\clang}{\textsf{Clang}\xspace}

\newcommand{\Gimli}{\textsf{Gimli}\xspace}
\newcommand{\HACL}{\textsf{HACL}$^\star$\xspace}
\newcommand{\fstar}{\textsf{F}$^\star$\xspace}

\newcommand*\figref[1]{Figure~\ref{fig:#1}}

\input{jasmin.lst}

\usepackage{textcomp}
\usepackage{listings}

\input{easycrypt.lst}

\lstnewenvironment{easycrypt}[2][]%
  {\lstset{language=easycrypt,caption={#2},#1}}%
  {}

\lstnewenvironment{easycryptfloat}[2][]%
  {\lstset{language=easycrypt,float=tp,caption={#2},#1}}%
  {}

\newcommand{\ecinput}[5][]%
{\lstinputlisting[language=easycrypt,linerange={#4},caption={#3},label={#5},#1]{#2}}

\newcommand{\ecinputfloat}[4]%
{\lstinputlisting[language=easycrypt,float=tp,linerange={#3},caption={#2},label={#4}]{#1}}

\newcommand*\ec[1]{{\footnotesize{#1}}}

\newcommand{\poly}{\textsf{Poly1305}\xspace}
\newcommand{\chacha}{\textsf{ChaCha20}\xspace}

\title{The Last Mile: High-Assurance and High-Speed Cryptographic Implementations}

\author{\IEEEauthorblockN{José Bacelar Almeida\IEEEauthorrefmark{1}, Manuel Barbosa\IEEEauthorrefmark{2}, Gilles Barthe\IEEEauthorrefmark{3}, Benjamin Grégoire\IEEEauthorrefmark{4}\\
    Adrien Koutsos\IEEEauthorrefmark{5}, Vincent Laporte\IEEEauthorrefmark{4},Tiago Oliveira\IEEEauthorrefmark{2}, Pierre-Yves Strub\IEEEauthorrefmark{6}}
  \IEEEauthorblockA{\IEEEauthorrefmark{1}University of Minho and INESC TEC%
  }
  \IEEEauthorblockA{\IEEEauthorrefmark{2}University of Porto (FCUP) and INESC TEC%
  }
  \IEEEauthorblockA{\IEEEauthorrefmark{3}MPI for Security and Privacy and IMDEA Software%
  }
  \IEEEauthorblockA{\IEEEauthorrefmark{4}Inria%
  }
  \IEEEauthorblockA{\IEEEauthorrefmark{5}LSV, CNRS, ENS Paris-Saclay%
  }
  \IEEEauthorblockA{\IEEEauthorrefmark{6}Ecole Polytechnique%
  }
}

\begin{document}

\maketitle

\begin{abstract}
  We develop a new approach for building cryptographic implementations.
  Our approach goes the last mile and delivers assembly code that is
  provably functionally correct, protected against side-channels, and as
  efficient as hand-written assembly. We illustrate our approach using
  \chacha-\poly, one of the mandatory ciphersuites in TLS 1.3, and
  deliver formally verified vectorized implementations which outperform
  the fastest non-verified code.

  We realize our approach by combining the \Jasmin framework, which
  offers in a single language features of high-level and low-level
  programming, and the \EasyCrypt proof assistant, which offers a
  versatile verification infrastructure that supports proofs of
  functional correctness and equivalence checking. Neither of these
  tools had been used for functional correctness before. Taken together,
  these infrastructures empower programmers to develop efficient and
  verified implementations by \lq\lq game hopping\rq\rq, starting from
  reference implementations that are proved functionally correct against
  a specification, and gradually introducing program optimizations that
  are proved correct by equivalence checking.

  We also make several contributions of independent interest, including
  a new and extensible verified compiler for \Jasmin, with a richer
  memory model and support for vectorized instructions, and a new
  embedding of \Jasmin in \EasyCrypt.

\end{abstract}

\section{Introduction}
\label{sec:intro}
Vulnerabilities in cryptographic libraries are challenging to detect
and/or eliminate using approaches based on testing or fuzzing. This
has motivated the use of formal verification for proving functional
correctness and side-channel protection for modern cryptographic
libraries~\cite{ZinzindohoueBPB17,Erbsen19,BondHKLLPRST17,FromherzGHPRS19}.
These approaches have been very successful, to the extent that some of
these verified libraries have been deployed in popular products like
Mozilla Firefox, Google Chrome, Android, etc. Despite their success,
these approaches compromise on efficiency, trust, and adoptability:

\begin{description}
\item [efficiency] Zinzindohoue \emph{et al}~\cite{ZinzindohoueBPB17}
  and Erbsen \emph{et al}~\cite{Erbsen19} produce verified C libraries
  that are as efficient as unverified C counterparts, for instance in
  the \openSSL library. However, many state-of-the-art cryptographic
  libraries are often written directly in assembly and outperform C
  implementations thanks to their use of vectorized instructions and
  clever platform- and algorithm-specific optimizations. Bond \emph{et
    al}~\cite{BondHKLLPRST17} verify assembly implementations but
  their implementations are not as fast as unverified assembly
  implementations;

\item [trust] it is theoretically possible to compile the verified C
  libraries from~\cite{ZinzindohoueBPB17,Erbsen19} with verified
  compilers, such as CompCert~\cite{Leroy06}, which provably preserves
  the functional correctness of programs.  However it is practically
  important to compile these libraries using a non-verified and
  aggressively optimizing compiler (e.g., \clang or \textsf{gcc}) in
  order to achieve reasonable levels of efficiency.  Therefore, the C
  compiler is implicitly assumed to preserve functional correctness of
  programs. Moreover, it is also implicitly assumed that compilation
  preserves protection against timing attacks. A similar issue arises
  with the work of Fromherz \emph{et al.}~\cite{FromherzGHPRS19}, who
  verify C implementations with inlined assembly.

\item [adoptability] the aforementioned approaches are primarily
  targetted to a posteriori verification of implementations produced
  by cryptographers, and do not aim to provide a practical framework
  that can be used by cryptographic engineers to simultaneously
  optimize and contribute to the formal verification of high-speed and
  high-assurance implementations.
\end{description}
In this paper we go the last mile: we develop a practical framework
that can be used by cryptographic engineers to build high-assurance
and high-speed implementations that are as fast or even faster than
their non-verified counterparts.

\subsubsection*{Methodology}
Our methodology allows developers to follow the typical optimization
process for low-level code.  We start from a readable reference
implementation, for which we prove functional correctness.
We then gradually transform the reference implementation into an
optimized (possibly platform-specific and vectorized) implementation,
and prove that each transformation preserves functional correctness.
This approach is similar to the ``game-hopping'' technique
used in provable security, except that we use it for functional
correctness of implementations rather than security of high-level
algorithms. 
In parallel, we check that implementations are safe using static 
analysis techniques, as the previous proofs are carried out in a simpler
semantics, which assumes that programs are safe, and compiler
correctness is generally stated for safe programs.
As a final step, we prove that that the optimized implementations 
correctly deploy mitigation against timing attacks: we adopt the
cryptographic constant-time approach~\cite{AlmeidaBBDE16}, and prove
that both the control flow and the memory accesses performed by the
optimized program are independent of secret values. 

This simple approach has important conceptual benefits. It emulates
the developer's mental process for writing optimized implementations,
and imposes a convenient separation of concerns between optimization
and verification. It also minimizes and helps structuring verification
work: functional correctness of the reference implementation is
established once and for all, even if the reference implementation is
used to derive several platform-specific implementations. Moreover,
correctness of transformations can be factored out using generic
lemmas.

\subsubsection*{Software infrastructure}
We realize our methodology using \Jasmin~\cite{CCS:ABBBGL17}, a
language and compiler for high-assurance and high-speed cryptography,
and \EasyCrypt~\cite{BartheGHB11,BartheDGKSS13}, a proof assistant for
provable security.
The \Jasmin language is designed to support \lq\lq assembly in the
head\rq\rq\ programming, i.e.\, it smoothly combines high-level
(structured control-flow, variables, etc.) and low-level (assembly
instructions, flag manipulation, etc.) constructs. This combination
makes it possible to program by \lq\lq game-hopping\rq\rq.
The \EasyCrypt proof assistant supports program logics for reasoning
about correctness and equivalence of imperative programs. It has been
used to mechanize \lq\lq game hopping\rq\rq\ security proofs for many
cryptographic schemes.
Neither \Jasmin nor \EasyCrypt has been used previously for proving
functional correctness of implementations. However, taken together,
they provide a convenient framework to develop efficient verified
implementations by \lq\lq game hopping\rq\rq.

We formally verify \Jasmin implementations. These implementations are
\emph{predictably} transformed into assembly programs by the \Jasmin
compiler. Predictability empowers \Jasmin programmers to develop
optimized implementations with essentially the same level of control
as if they were using assembly or domain-specific languages such as
\textsf{qhasm}. Moreover the compiler is verified (in the \Coq proof
assistant) thus guarantees are carried to assembly code.

\subsubsection*{Technical contributions}
We build on \Jasmin and \EasyCrypt to instantiate a new methodology 
for the development of high-speed and high-assurance crypto code and
demonstrate the resulting framerwork by giving new, fully verified, 
assembly implementations of standard cryptographic algorithms that are 
fastest than their best known (non-verified) counterparts.
In detail, our technical contributions are as follows:
\begin{enumerate}[leftmargin=*]
\item We enhance the \Jasmin framework with a richer memory model,
  supporting values of different sizes, several language extensions,
  including intrinsics for vectorized instructions, and a new compiler
  design that favors extensibility. We leverage these enhancements and
  use our ``game hopping'' approach to obtain highly optimized
  vectorized implementations for \chacha, \poly and \Gimli;

\item We implement an embedding of \Jasmin programs in the \EasyCrypt
  proof assistant~\cite{BartheDGKSS13}. The embedding naturally
  supports proofs of functional equivalence and functional
  correctness. We also develop a variant of the embedding to support
  automatic proofs of protection against side-channel attacks,
  concretely that control flow and memory accesses are independent
  from secret inputs (aka.\, cryptographic constant-time);
 
\item We prove functional correctness of reference implementations,
  and equivalence between reference and optimized implementations for
  \chacha, \poly and \Gimli. Statements of functional correctness for
  the first two primitives are taken from, or given in a style similar
  to, \HACL, to guarantee formal interoperability.  In the case of
  \Gimli we show how to use a readable \Jasmin reference
  implementation, which can be syntactically very close to the
  specification given in cryptographic standards, as a goal for
  proving functional correctness of optimized code.
\end{enumerate}

\medskip
We note that a crucial factor in achieving these results is our ability
to tame the verification effort by relying on equivalence checking.
Due to the characteristics of \Jasmin, the verification workload for the 
reference implementations is comparable (or even smaller, due to the less
elaborate memory model) to that of carrying out formal verification of
C code. 
The power of the relational reasoning offered by \EasyCrypt permits
bridging reference implementations and optimized implementations with
relatively low effort, and the automation offered by the tool suffices to
deal with proof goals for side-channel protection.

\subsubsection*{Limitations}
We have only demonstrated our approach for selected primitives and x86
platforms. However, we plan to exploit extensibility of our compiler
to support ARM platforms. Moreover, we plan to cover a broader range
of primitives, and note that compatibility with \HACL specifications
would make it possible to use \Jasmin implementations as drop off
replacements of \HACL implementations for \chacha and \poly.

\subsubsection*{Software and proofs}
Available at \url{https://github.com/jasmin-lang/jasmin} and
\url{https://github.com/tfaoliveira/libjc}

\subsubsection*{Outline}
In the next section we use an example to illustrate our
methodology. In Section~\ref{sec:langcomp} we describe our
extensions to \Jasmin and in Section~\ref{sec:eqcheck}
we describe the supporting development in \EasyCrypt.
Then, in Sections~\ref{sec:casestudies} and~\ref{sec:bench}
we describe our other case studies and provide a thorough
performance evaluation of our code in comparison to 
alternative implementations. Related work is reviewed in
Section~\ref{sec:related} and concluding remarks appear 
in Section~\ref{sec:conclusion}.













\section{Motivating example: \poly}
\label{sec:motivex}

We illustrate our methodology using \poly~\cite{Bernstein05}, an
authentication algorithm that is used together with \chacha as one of the two
mandatory ciphersuites for TLS 1.3.
\poly is a one-time authenticator (the key should only be used
once) that allows the sender to attach a cryptographic tag $t$ 
to a transmitted message $m$. The receiver of the message
should be able to derive the same session key $k$ autonomously, and
recompute the tag on the received message. If the tags match, 
the receiver is assured that only the sender could have 
transmitted it, provided $k$ is secret and authentic.

\subsubsection*{Algorithm overview}

\poly takes a 32-byte one-time key $k$ and a
message $m$ and it produces a 16-byte tag $t$.  The key $k$ is seen as
a pair $(r,s)$, in which each component is treated as a 16-octet
little-endian number, with the following format restrictions: octets
$r[3]$, $r[7]$, $r[11]$ and $r[15]$ should have their top 4 bits
cleared, whereas octets $r[4]$, $r[8]$ and $r[12]$ are required to
have their two lower bits cleared.  For the purpose of this paper we
will assume that $k=(r,s)$ is generated as a pseudorandom 256-bit
string, after which $r$ is \emph{clamped} to its correct format.

To authenticate a message $m$, it is split into 16-byte blocks
$m_i$, for $i \in [1,2,\dots]$.  Each block $m_i$ is then converted
into a 129-bit number $b_i$ by reading it as a 16-byte little-endian
value and then setting the 129-th bit to one (the last block is treated
differently). The authenticator $t$ is computed by sequentially
accumulating each such number into an initial state $a_0=0$ according
to the following formula: $a_{i} = (a_{i-1} + b_i) \times r \pmod p $,
for $i \in [1,2,\dots]$ and where $p = 2^{130}-5$ is prime.  Finally, the
secret key $s$ is added to the accumulator (over the integers) 
and the tag $t$ is simply the lowest 128 bits of the result serialized 
in little-endian order.
The choice of $p$ is crucial for optimization, as it is close to a
power of $2$: modular reduction can be performed by
first reducing modulo \(2^{130}\) and then adjusting the result using
a simple computation that depends on the offset $5$.

\subsubsection*{Specification}
Our goal is to prove that our optimized implementation of \poly is
functionally correct with respect to the high-level specification
presented in Figure~\ref{fig:polyspec}. The specification is written
in \EasyCrypt and matches the \HACL specification for \poly
in that the computation of the tag is expressed as the following
functional operators, which carry out a fold over a list of
values in $\mathbb{Z}_p$. 
\begin{lstlisting}[language=easycrypt,xleftmargin=0pt,xrightmargin=0pt,basicstyle=\footnotesize,frame=none, escapeinside={/*}{*/}]
op poly1305_loop (r : zp) (m : Zp_msg) (n : int) =
  foldl (fun h i => (h + nth m i) * r) /* $0_{\mathbb{Z}_p}$ */ [0;...;n-1].

op poly1305_ref (r : zp) (s : int) (m : Zp_msg) =
  let h' = poly1305_loop r m (size m) in  
  (((asint h') % /* $2^{128}$ */) + s) % /* $2^{128}$ */ .
\end{lstlisting}
\begin{figure}[t]
\framebox{
  \begin{minipage}{.45\textwidth}
    \input{listings/poly1305_spec}
  \end{minipage}}
\caption{\poly specification in \EasyCrypt.}\label{fig:polyspec}
\end{figure}
This specification is used to express the following correctness contract 
over the execution of our implementations:
\begin{lstlisting}[language=easycrypt,xleftmargin=0pt,xrightmargin=0pt,basicstyle=\footnotesize,frame=none]
      Glob.mem = mem /\ args = (out,inn,inl,k) /\
        poly1305_pre r s m mem inn inl k ==> 
      poly1305_post mem Glob.mem out r s m
\end{lstlisting}

The contract relies on an axiomatic model of the \Jasmin language
semantics that has been created in \EasyCrypt.
In this particular case, the contract imposes that the memory 
\ec{Glob.mem} in the final state is identical to the initial memory, 
except for the fact that it now encodes 
the correct authenticator at position \ec{out}. Correctness is defined
with respect to the values stored in the initial memory, whose contents
are interpreted (according to the encoding rules of \poly) as containing 
a message \ec{m} of length \ec{inl} bytes stored at position \ec{inn}, and a 
key with components \ec{r} and \ec{s} stored at position \ec{k}.

In detail, the precondition and post-condition are shown in
Figure~\ref{fig:polyspec}.
The pre-condition requires the implementation
to correctly \emph{lift} the message encoded in memory to some 
representation of $\mathbb{Z}_p$, tweaking the necessary bits as specified
by \poly, as defined by  operators \ec{load\_block}
and \ec{load\_clamp}. We illustrate the latter:
\begin{lstlisting}[language=easycrypt,xleftmargin=0pt,xrightmargin=0pt,basicstyle=\footnotesize,frame=none]
 op load_clamp(mem: global_mem_t) (ptr : address) = 
    let x = loadW128 mem ptr in
    let xclamp = 
       x & (W128.of_int 0xFFFFFFC0FFFFFFC0FFFFFFC0FFFFFFF) in
    Zp.inzp (W128.to_uint xclamp).
\end{lstlisting}
Note that the lifting to $\mathbb{Z}_p$ is represented 
by operators \ec{inzp} and \ec{asint}. 
Conversely, the post-condition requires the implementation to 
correctly encode the final tag (a multi-precision integer) back into memory.

\subsubsection*{Implementations}
We reach our optimized code from the specification
through a sequence of implementations: 
\begin{description}[leftmargin=*]
\item [abstract implementations:] we first create an imperative
  version of the specification (see Figure~\ref{fig:polyref}) that
  relies on computations over the abstract type \ec{zp}. We then apply a
  series of transformations (including loop transformations and code
  modularization using inlineable functions) to obtain a code that
  approximates the control flow from Figure~\ref{fig:polystruct},
  and which we explain below.
  
\item [reference implementations:] we replace computations in \ec{zp}
  with calls to functions that deal with explicit representations of
  values in \ec{zp}. Intuitively, this program corresponds to a
  \Jasmin program, whereas the abstract implementations are just
  \EasyCrypt programs which we use as proof artifacts.
  However, this reference implementation is not yet fully optimized.

\item [optimized implementations:] we apply a series of transformations
  to restructure the reference implementation in a code that exhibits
  parallelism and replaces sequential code by vectorized instructions.
  The final, fully optimized, implementation in the sequence is
  described next.
\end{description}

\subsubsection*{High-speed high-assurance implementation}
\begin{figure}
\centering
\includegraphics[width=.5\textwidth]{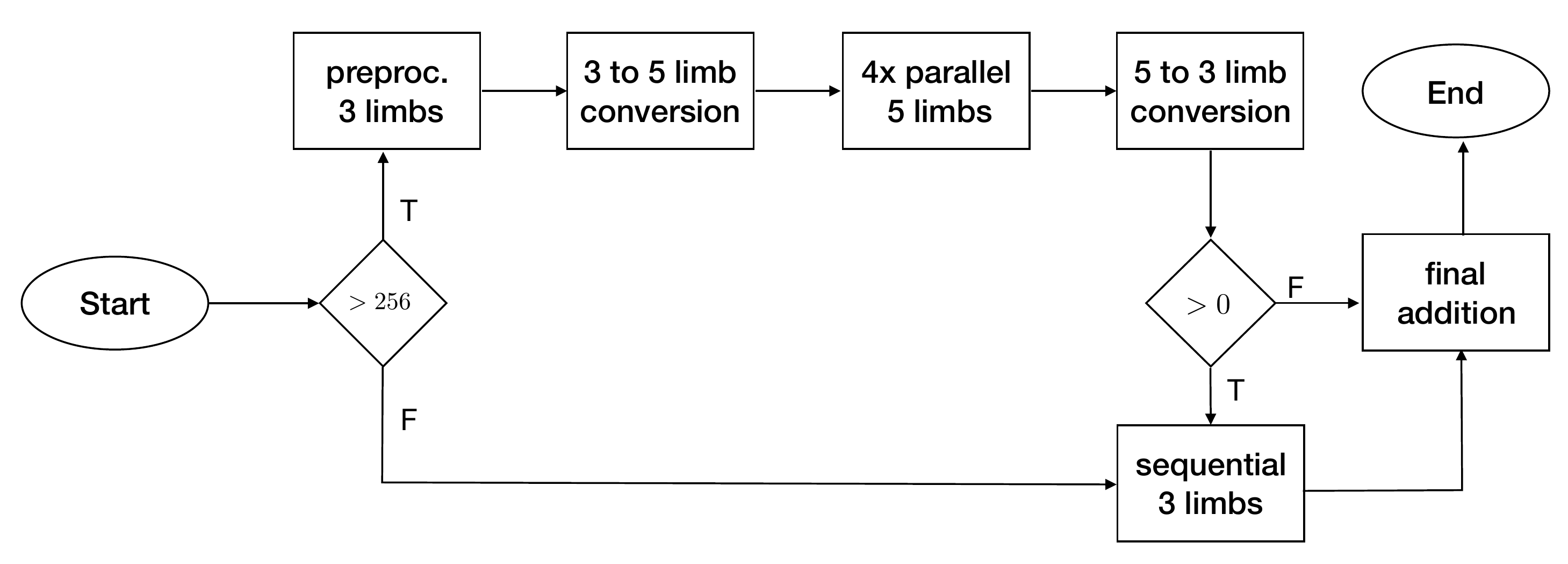}
\caption{Structure the optimized \poly implementation.}
\label{fig:polystruct}
\end{figure}

Our fully optimized code takes advantage of the assembly in the
head style of programming supported by \Jasmin.  We rely on the
high-level constructs in \Jasmin to deploy mixed representation
optimizations, which combine sequential and parallel processing as
shown in Figure~\ref{fig:polystruct}.

The implementation first checks whether we are dealing with a small or
large message (over 256 bytes). For small messages, 
it calculates the tag by representing values in
$\mathbb{Z}_p$ packed into three 64-bit words (the most significant
word for a residue will only use $2$ bits).  For large messages, a
mixed representation computation is used.  First, some precomputation
necessary for parallel calculations is performed using the packed
representation; then the values are converted to a 5-limb
representation using radix $2^{26}$ stored into five 64-bit words.
This leaves room in each word so that limb-wise multiplication can be
performed safely in 64-bit architectures, as well as accumulating
multiple additive carry operations.  Parallel computation of 4 message
blocks at a time is then implemented using vectorized operations over
this representation.  Finally, the result is converted back to the
packed representation and any remaining message blocks are processed
as for short messages.

 The high-level control flow and
(inlineable) function modularization of \Jasmin are crucial to allow
managing the code complexity, whereas the low-level features permit
controlling instruction selection and scheduling in order to fine-tune
performance.  An example of our use of the low-level features of the
language is given in Figure~\ref{fig:polyopt}.  The optimized
implementation relies on {\sc avx2} {\sc simd} instructions, for which \Jasmin
provides syntactic sugar: shift and add operators are annotated with
type information (\ec{4u64}) indicating that the selected instructions
act on $4$ unsigned $64$-bit words in parallel.  Indeed, this code
snippet is a part of the parallelized $5$-limb implementation, as can
be seen by the type of the input \ec{x}, which contains $4$ values in
$\mathbb{Z}_p$, each represented using $5$-limbs. All of these values
are processed in the same way using the \textsc{simd} instructions.

\begin{figure}[t]
\begin{jasmincode}\input{listings/poly1305_opt.jazz}\end{jasmincode}
\caption{Example of optimized \Jasmin{} low-level code.}\label{fig:polyopt}
\end{figure}

\subsubsection*{Correctness of abstract and reference implementations}
The proof of the baseline abstract implementation with respect to the
functional specification uses standard Hoare logic. The proof of the
remaining abstract implementations is carried by \lq\lq game
hopping\rq\rq, i.e.\, each step is justified using relational Hoare
logic.

The proof of equivalence from the baseline abstract implementation
uses a series of functional correctness lemmas that modularize the
computation of each operation in \ec{zp} in the two representations
described above, including arithmetic, conversion and load/store
operations.

 \begin{figure}[t]
\framebox{
  \begin{minipage}{.48\textwidth}
    \input{listings/poly1305_ref}
\end{minipage}}
\caption{Imperative reference specification in \EasyCrypt.}
\label{fig:polyref}
\end{figure}

 \medskip
Figure~\ref{fig:zpcontract} shows one such auxiliary lemma for the
conversion of the results of the $4$ parallel computations over the
$5$-limb representation to a single value over $3$-limb
representation.  Unlike the rest of the correctness proof, which
requires minimal effort, the proofs of these lemmas require ingenuity
and user interaction. This is because the proofs of these auxiliary
lemmas use algebraic reasoning similar to proofs of other
multi-precision computations that are common in
cryptography~\cite{CCS:CHLSTW14}. These proofs are reusable and they
could be partially automated (see Sections~\ref{sec:eqcheck}
and~\ref{sec:related}).

\medskip

We note that this strategy of performing several hops with abstract
implementations considerably simplifies the equivalence proofs.

\begin{figure}[t]
\framebox{
  \begin{minipage}{.48\textwidth}
    \input{listings/zpcontract}
\end{minipage}}
\caption{Example of representation correctness lemma. In addition
to functional correctness, the contract binds the implementation
to using only $4$ bits on the most significant output limb, assuming
that all the input limbs are using at most $27$-bits.}
\label{fig:zpcontract}
\end{figure}

\subsubsection*{Equivalence checking the vectorized implementation}
At this point in the proof, all the non-vectorized code in the reference
implementation matches the code in the extracted \Jasmin implementation.
The final part of the proof connects our reference implementation
to the fully optimized code via three hops that rely on two 
new dedicated \EasyCrypt features:
\begin{itemize}[leftmargin=*]
\item The first hop rearranges code so that code corresponding
to each single-instruction-multiple-data ({\sc simd}) instruction is
modularized as a call to a procedure in a special \EasyCrypt module
called \ec{Ops}. Here we take advantage of a new \EasyCrypt
meta-tactic that permits rearranging $n$ repetitions
of the same sequence of $k$ instructions into a 
sequence of $k$ blocks, each with $n$ identical instructions (intuitively
each of these blocks corresponds to a {\sc simd} instruction).
\item The second hop replaces calls to \ec{Ops} with calls to
a different module \ec{OpsV}, where the content of each procedure
now makes a single call to the corresponding {\sc simd} instruction
(see Figure~\ref{fig:opsvops} for an illustrative example). 
This hop is justified with a once-and-for-all proof that, 
using the axiomatic semantics of \Jasmin 
expressed in \EasyCrypt, establishes the functional equivalence
of the \ec{Ops} and \ec{OpsV} modules.
\item The final hop simply shows that the \EasyCrypt implementation
obtained from the previous transformation is equivalent to the
extracted optimized \Jasmin code.
\end{itemize}

\begin{figure}[t]
\framebox{
\begin{minipage}{.48\textwidth}
\input{listings/opsvops}
\end{minipage}}
\caption{\ec{Ops} and \ec{VOps} modules.}
\label{fig:opsvops}
\end{figure}

In Section~\ref{sec:casestudies} we report on other equivalence
proofs we have completed and also on the performance enhancements
obtained via the associated optimizations.

\subsubsection*{Protection against timing attacks}
Low-level cryptographic software implementations are expected to
satisfy a security-critical non-functional property commonly known as
{\em constant-time}, as mitigation against timing attacks. 
This mitigation consists of the following two restrictions:
i. all branching operations depend only on public
values (here {\em public} is defined by explicitly identifying which
parts of the initial state can influence the control-flow); and
ii. the memory addresses that are accessed by the implementations
depend only on public values. Intuitively, these restrictions combined
with mild assumptions on the underlying hardware processor guarantee
that the execution time of the program (even accounting for
micro-architectural features like cache memories) will be fixed once
the public part of the input is fixed, ruling out timing attacks.

It is well-known that mitigation against timing attacks 
can be modeled as observational non-interference.
 That is, one can define an instrumented semantics,
where a distinguished leakage variable (modeled as a list of events)
records execution of time-varying instructions, targets of conditional
jumps and memory accesses. Then, a program is secure iff every
two executions with possibly different secret inputs (but equal public
inputs) yield equal leakage.

We define an embedding of the instrumented semantics of \Jasmin
programs in \EasyCrypt and use this embedding to translate our
optimized implementation of \poly. The proof that a program
correctly implements mitigation against timing attacks is 
carried with minimal user interaction using existing tactics 
for relational Hoare logic.

\subsubsection*{Guarantees over assembly code}
The \Jasmin compiler is formally verified for functional correctness (in \Coq),
therefore we know that safe \Jasmin source programs are compiled into
safe and functionally equivalent assembly programs---we discuss safety
in detail in Section~\ref{sec:langcomp}. In addition, Almeida \emph{et
  al}~\cite{CCS:ABBBGL17} informally argue that the \Jasmin compiler
preserves mitigations against timing attacks in the constant-time model.  
This informal claim has been reinforced by a
formal proof that many optimization passes preserve such mitigations~\cite{BartheGL18}, 
although the connection between these two works has not been established yet.

\subsubsection*{Efficiency}
\HACL contains a verified C implementation of \poly.
Vale~\cite{BondHKLLPRST17} and Vale/F*~\cite{FromherzGHPRS19} also
prove functional correctness of the assembly implementation of \poly
for 64-bit operations. In section Section~\ref{sec:bench} we compare our
code with these and other implementations and show that it
outperforms the best non-verified code for \poly. 

We note that there is no magical way of improving performance and we
do not claim to have a way to replace the expert programmer. 
Indeed, we apply the same optimization ideas used by the best 
implementations, namely those adopted in \textsf{OpenSSL} and~\cite{Goll2015VectorizationOP}.
What we {\em do} claim is that we have a tool-supported methodology
that allows us to bring the expert programmer on-board in the process
of verifying functional correctness, in that many of the hand-crafted
optimization steps can be justified with equivalence-checking proofs 
that are simple to carry out with the developers' input. This includes
vectorization and instruction scheduling as prominent examples.
Furthermore, the more challenging parts of the functional correctness
proofs are not harder (one could even argue that they are simpler due
to the memory model of Jasmin) that proofs carried out over high-level
languages such as C.

\section{\Jasmin language, compiler and analyzer}\label{sec:langcomp}
In this section we describe the improvements to \Jasmin required 
for writing our highly-optimized fully-verified implementations. 
This includes language extensions and a significant
refactoring of the infrastructure and compiler.
In Appendix~\ref{app:jasmin} we give a summary of the original
key design choices for \Jasmin.
Here we start by listing the main challenges we addressed
in improving previous work \cite{CCS:ABBBGL17}. 

The original \Jasmin framework~\cite{CCS:ABBBGL17} has been used for
writing high-speed cryptographic code. Unfortunately, this framework
exhibits three key limitations:
\begin{itemize}[leftmargin=*]
\item simplified memory model: only 64-bit values were supported in
  the original \Coq formalization, which excluded implementations that
  rely, for example, on byte-level access to memory or different views
  of arrays, and vector instructions;

\item monolithic design: the original \Coq formalization was developed
  in the prevailing style for verified compilers, and not intended to
  be easily extensible;

\item basic proof infrastructure: the original implementation featured
  a verification condition generator (the standard tool for deductive
  verification) and did not support the process of writing optimized
  implementations described in the previous section.
\end{itemize}
We address these issues as follows. First, we develop a new memory
model to support value of any width and idioms for efficient low-level
code. Second, we propose a mechanism, called instruction descriptor,
which incorporates the information required to handle the instruction
from source to assembly, and implement the language and compiler based
on instruction descriptors. Third, we develop a verification
infrastructure that supports proofs by game hopping. We develop the
first two points in this section, and the third point in the next
section.

\subsection{Compiler design}
Compilers, and in particular verified compilers, are typically written
for well-defined source languages and architectures. Moreover, it is
generally assumed, at least implicitly, that compiler extensions will
be developed by compiler writers. This is perfectly reasonable, but
has concrete practical implications: any extension or modification to
the compiler requires multiple modifications spread over the codebase
of the compiler. Furthermore, in the case of verified compilers, it is
also necessary to perform multiple modifications across the compiler
proof.

In our context, this status quo is unsatisfactory for two reasons:
first, the source language is not as closed as traditional languages:
in particular, it is designed to support assembly in the head and to
grow organically to support richer sets of instructions and eventually
multiple platforms. Ideally, these extensions should be manageable by
external contributors with limited skills in formal verification.
Therefore, we introduce the notion of \emph{instruction descriptor},
which packs all the knowledge and the proof obligations required for
adding new instructions to the compiler. The use of descriptors makes
the compiler more easily extensible. In addition, the approach is
generic and applicable to other compilers.

Machine instructions can be made available as \emph{intrinsic}
operators at the source level.  Due to historical reasons and
micro-architectural constraints, each instruction has a specific
“calling-convention”.  For instance, many instructions implicitly
write part of their output to the \emph{flags} register.  Also,
several instructions operate \emph{in place}: their main destination
should be the same as one of their sources.  These constraints are
irrelevant to program verification and enforcing them early in the
compilation process would not bring any benefit.  As an example, the
x86 \textsf{MUL} instruction takes two inputs: one explicitly and the
other one implicitly from the \textsf{RAX} register; its output is
always written to the \textsf{RDX}, \textsf{RAX} and flags registers.
On the opposite, these operators are uniformly described as pure
functions at the source level.  In the case of the \textsf{\#x86\_MUL}
intrinsic, it takes two values as input and produces seven values as
output.  Therefore, the compilation of these simple instructions is
not trivial: on one hand register allocation must enforce the various
architectural constraints; on the other hand the generation of
assembly code should check that it is safe to move from a functional
semantics to an imperative one with side effects.  The associated
correctness proof is also tedious and slightly involved.

Therefore, we have built a novel verified compiler infrastructure so
that new instruction can be added by constructing a descriptor and
adding it to the development. This permits automatically implementing
compilation and extending the correctness proof to support the new
instruction.  In particular, descriptors permit extracting rules for
register allocation, as they include meta-data about which registers
are read or written by an assembly instruction, which instruction
performs memory accesses, etc.  This information is also useful to
extend support for verification of non-functional properties, in our
case mitigation of timing attacks.

For instance, the descriptor for the \textsf{\#x86\_MUL sz}
(non-truncated unsigned multiplication of words of \textsf{sz} bits)
is constructed from the following data:
\begin{itemize}
\item\textsf{[:: F OF; F CF; F SF; F PF; F ZF; R RDX; R RAX]}
  the list of destinations where to write the output values,
  here implicitly to some flag registers and to the \textsf{RDX} and \textsf{RAX} registers;
\item\textsf{[:: R RAX; E sz 0]}
  the list of sources from where to read the input values,
  here implicitly from the \textsf{RAX} register and from the first argument;
\item\textsf{[:: TYoprd]}
  the type of the emitted instruction,
  here it has one operand;
\item\textsf{(MUL sz)}
  the instruction to emit.
\end{itemize}
The descriptor also contains a few well-formedness arguments (proved
by computation) and a correctness argument which links the high-level
functional semantics of the intrinsic operator to the low-level
imperative semantics of the machine instruction.

As an additional advantage of this design, pseudo-instructions can be
seamlessly introduced.  For instance, a common pattern for
zero-initializing a register is to \textsc{xor} it with itself.
However in the \Jasmin{} source language, uninitialized registers have
undefined values that should not be used in computations.  We have thus
added an operator \textsf{set0} which takes no input and returns a
zero value; its descriptor maps it to the \textsf{r \(\mapsto\)
  XOR~r~r} instruction which takes only one argument: its destination.

\subsection{\Jasmin language and memory model}
\subsubsection*{Memory model}
Although x86\_64 microprocessors mainly provide 64-bit registers,
programs may manipulate values of various bit widths.  Ultimately,
there are only registers of 64 bits. Values of smaller sizes do fit,
but some bits are undefined.  If the value is larger than what is
expected by an operation, this is not an issue: only the relevant bits
are used.  The behavior when writing a small value into a register
depends on the operation and its size: the high bits of the
destination registers may be preserved or zeroed.  The exact behavior
at the micro-architectural level is very intricate: it would be unwise
to expose it to the programmer.  Therefore, we define two semantics:
a source semantics that is uniform and convenient for reasoning; and
a compiler semantics, in which variables may hold ``partial'' values,
i.e., with some of their most significant bits being undefined.

In any case, operators at some size may be applied to arguments of a
larger size: arguments are implicitly truncated.  The compiler needs
not do anything special, since this is the semantics at the assembly
level: operators extract the relevant part from their register
operands.  Technically, we require that functions have a signature and
every assignment be decorated with the type of the assigned value.  By
analogy to an identity operator, assignment truncates the value to the
given type, which enables to soundly compile copies using a
(truncating) \textsf{mov} instructions.

Another extension that permits dealing with varying length operations
is a refinement of the memory model specification to allow
``type-punning'' — reading and writing distinct but overlapping ranges
of addresses.  Interestingly, the precise behavior need not be
specified in order to prove the correctness and security of the
compilation.  Note however, that to reason about \Jasmin programs
relying on such memory access patterns, the instance of the memory
model might need to be refined.

\subsubsection*{Flexible views of stack arrays} 
To efficiently implement functions like the C function \textit{memcpy}, 
or the processing of the last (potentially incomplete) block of plaintext
in{} a stream cipher, it is important to be able to use the same pointer to
read and store data of varying sizes.
For memory accesses, this follows a direct consequence of supporting 
various sizes in \Jasmin. However, the stack in \Jasmin is not seen
as addressable memory at the source level, although stack arrays
are compiled as pointers into the stack.
To allow the same flexibility in stack operations, we have added a 
special feature for arrays in the stack, which allows reading/writing 
words of different sizes, as illustrated in Figure~\ref{fig:load_last_add}.

\begin{figure}
\begin{jasmincode}\input{listings/load_last_add.jazz}\end{jasmincode}
\caption{\label{fig:load_last_add}Load and add the final bytes}
\vspace{-12pt}
\end{figure}

As can be seen in the figure, a stack array can be seen as a contiguous 
sequence of bytes, which is very convenient when only a part of the
array ends up being used.
Aliasing and overlapping accesses issue may thus arise, but they are 
scoped to a single array: at the source level, stack arrays as a whole 
enjoy a value semantics, are disjoint, etc.

\subsubsection*{Vector instructions}
Instruction descriptors and our more general memory model allow us to
integrate vector instructions in the \Jasmin language. Following the
design principles of \Jasmin, we consider both generic, zero-cost,
portable and non-portable instructions.

One example of portable instructions is parallel addition.  For these
operations, the language provides a convenient syntax.  For instance,
the instruction \textsf{x +\jasmintype{4u64}= z;} in which variables
\textsf{x} and \textsf{z} have type \textsf{\jasmintype{u256}},
performs four parallel 64-bit additions on vectors \textsf{x} and
\textsf{z} and assigns the result to variable~\textsf{x}.

Non-portable instructions are available through the general syntax for
\emph{intrinsics}, e.g., \textsf{\#x86\_VPSHUFD\_256}.  This operator
shuffles the four 64-bit elements of its first argument.  The exact
shuffling is specified by its second argument, which is an 8-bit
value.  This value is best seen as a vector of four 2-bit numbers
describing, for each element of the destination vector, its original
position in the source vector.  We have introduced a convenient syntax
to represent this kind of constant values.  For example,
\figref{chacha20_shuffle} shows a shuffling routine that is part of
the \chacha operation.

\begin{figure}
\centering
\begin{jasmincode}
\jasminkw{fn} \jasmindname{shuffle_state}(\jasminstorageclass{reg} \jasmintype{u256}[4] k) \jasminarrow{} \jasminstorageclass{reg} \jasmintype{u256}[4] \{\\
\jasminindent{1}k[1] = \#x86_VPSHUFD_256(k[1], (\jasmintype{4u2})[ 0, 3, 2, 1]);\\
\jasminindent{1}k[2] = \#x86_VPSHUFD_256(k[2], (\jasmintype{4u2})[ 1, 0, 3, 2]);\\
\jasminindent{1}k[3] = \#x86_VPSHUFD_256(k[3], (\jasmintype{4u2})[ 2, 1, 0, 3]);\\
\jasminindent{1}\jasminkw{return} k;\\
\}
\end{jasmincode}
\caption{\label{fig:chacha20_shuffle}Shuffling function of \chacha{}}
\end{figure}

\begin{figure}
  \begin{jasmincode}\input{listings/bash_R4.jazz}\end{jasmincode}
  \caption{\label{fig:bash_R4}Parallel rotation function}
\end{figure}

\subsubsection*{Dynamic globals}
The initial version of \Jasmin allowed parameters in source code: constants
that are inlined very early in the compilation process similarly to C macros.
However, not all constants are the same and, for performance reasons, some
constants are best stored in the code segment, e.g., to take advantage of
\textsc{rip}-based addressing.\footnote{In this mode,
the data is stored within the code segment and referenced through a small offset
relative to the current value of the instruction pointer.}

To permit taking advantage of these features, our extension to \Jasmin
permits tagging local variables as \textsf{\jasminstorageclass{global}}.
These will behave as any other local variable,
but will be compiled to a code-segment constant value.
For this to be possible, their value should be known at compile-time,
after expansion of parameters, function call inlining, loop unrolling and
constant propagation.
The compiler will ensure that globals with equal values are merged.

This is a very useful mechanism, when an immediate argument to an instruction
is best described by a computation, as in vector instructions in which the immediate
value describes a permutation, or a vector of shift counts.
As an example, the \jasmindname{R4} function shown in \figref{bash_R4} performs 
four parallel rotations on a vector of 64-bit values.
The first argument is a vector of inline values that correspond to the bit counts 
of these different rotations.
These rotations are implemented using one left shift by the given bit count,
one right shift by the complementary bit count,
and a final \textsc{xor} of the results of the two shifts.
The bit counts are computed as inline values and stored in the code segment.

\subsection{Compiler correctness and safety analysis}
We have formalized the operational semantics of \Jasmin programs and
x86 assembly code in the \Coq proof assistant. The formalization is
based on the new memory model, and supports instruction extensions,
including {\sc simd}. We have also developed an extensible compiler
architecture based on instruction descriptors, and proved that the
compiler is correct. This means that the result of the compilation
preserves the semantics of the original \Jasmin program, assuming that
the program is well-typed, safe, terminating, and accepted by the
compiler---the compiler may still fail for well-typed and safe
programs, for instance because the compiler does not perform spilling.

We have also extended the \Jasmin compiler to verify that the source
program is safe, using a fully automated static analyser, as well as
terminating, using a simple analysis based on ranking functions.
Concretely, for safety we check for the absence of division by zero,
out-of-bound array accesses and variable initialization.  Moreover, we
need to ensure that, during the execution of the \Jasmin program, all
loads and stores take place in allocated chunks of the memory (i.e. a
specification of valid memory regions, which define the memory calling
contract).  We do not require the user to supply the static analyser
with the allocated memory ranges. Instead, we automatically compute an
over-approximation of the offsets that must be allocated in the
memory. Once the analysis is complete, the user is notified of the
inferred ranges, which are sufficient conditions under which the
program is safe. Since the offsets accessed in the memory may depend
on the inputs of the program, these are symbolic conditions involving
the initial value of the inputs. We consider polyhedral conditions,
i.e. conjunctions of linear inequalities. For example, in the case of
\poly, we automatically infer the following ranges:
\begin{alignat*}{4}
  &\textsf{range}(\textsf{out})&:\;\;& \textsf{out} + [0; 16[
  \qquad&&\textsf{range}(\textsf{inlen})&:\;\;& \emptyset\\
  &\textsf{range}(\textsf{k})&:\;\;& \textsf{k} + [0; 32[
  \qquad&&\textsf{range}(\textsf{in})&:\;\;& \textsf{in} + [0;\textsf{inlen}[
\end{alignat*}
Our analysis is based on abstract interpretation
techniques~\cite{DBLP:conf/popl/CousotC77}, and uses the
Apron~\cite{DBLP:conf/cav/JeannetM09} library of numerical domains. To
over-approximate the memory accesses, we use a symbolic {\em points-to}
abstraction combined with the polyhedra domain. Operations in
the polyhedra domain have a worst-case exponential complexity in the
number of variables. Therefore, we perform a pre-analysis to detect
which variables must be included in the relational domain. Moreover,
we allow to user to help the analysis by indicating which input
variables are pointers ($\textsf{k}$,$\textsf{in}$ and $\textsf{out}$
in \poly), and which variables must be included in the relational
domain ($\mathsf{inlen}$ in \poly).




\section{Source-level verification}
\label{sec:eqcheck}

This section describes our embedding of \Jasmin in \EasyCrypt. We use
this embedding for proving correctness of reference implementations,
equivalence between reference and optimized implementations, and
finally correct mitigation of timing attacks.

\subsection{Overview of \EasyCrypt}
\EasyCrypt~\cite{BartheDGKSS13} is a general-purpose proof assistant
for proving properties of probabilistic computations with adversarial
code. It has been used for proving security of several primitives and
protocols~\cite{BartheGLB11,BartheGHB11,BartheCLS15,AlmeidaBBDGLP17}.

\EasyCrypt implements program logics for proving properties of
imperative programs. In contrast to common practices (which use
shallow or deep embeddings), the language and program logics are
hard-coded in \EasyCrypt---and thus belong to the Trusted Computing
Base. The main program logics of \EasyCrypt are Hoare logic, and
relational Hoare logics---both operate on probabilistic programs but
we only used their deterministic fragments. The relational Hoare logic
allows to relate two programs, possibly with very different control
flow. In particular, the rule for loops allows to relate loops that do
not do the same number of iterations.  This is essential for proving
correctness of optimizations based on vectorization, or when the
optimization depends the input message length.

The program logics are embedded in a higher-order logic which can be
used to formalize and reason about mathematical objects used in
cryptographic schemes and also to carry meta-reasoning about
statements of the program logic. Automation of the ambient logic is
achieved using multiple tools, including custom tactics (e.g.\, to
reason about polynomial equalities) and back-end to SMT solvers.  For
the purpose of this work, we have found it convenient to add support
for proof by computation. This tool allows users to perform proofs
simply by (automatically) rewriting expressions into canonical forms.

\subsection{Design choices and issues}
Rather than building a verified verification infrastructure on top of
the \Coq formalization of the language (a la VST~\cite{Appel14}), we
opt for embedding \Jasmin into \EasyCrypt. We choose this route for
pragmatic reasons: \EasyCrypt already provides infrastructure for
functional correctness and relational proofs and achieves reasonable
levels of automation. On the other hand, embedding \Jasmin in
\EasyCrypt leads to duplicate work, since we must define an embedding
of the \Jasmin language into \EasyCrypt. Although we already have an
encoding of \Jasmin into \Coq, we cannot reuse this encoding for two
reasons: first, we intend to exploit maximally the verification
infrastructure of \EasyCrypt, so the encoding should be fine-tuned to
achieve this goal. Second, the \Coq encoding uses dependent types,
which are not available in \EasyCrypt. However, these are relatively
simple issues to resolve, and the amount of duplicate work is largely
compensated by the gains of using \EasyCrypt for program verification
(also note that building a verified verification infrastructure in
\Coq requires some effort).

\subsection{Embedding \Jasmin in \EasyCrypt}
The native language of \EasyCrypt provides control-flow structures
that perfectly match those in \Jasmin, including {\sf if}, {\sf while}
and {\sf call} commands. This leaves us with two issues: 1) to encode
the semantics of all x86 instructions (including SIMD) in \EasyCrypt;
and 2) to encode the memory model of \Jasmin in \EasyCrypt.

\subsubsection*{Instruction semantics}
Our formalization of x86 instructions aims at being both readable and
amenable to building a library of reusable properties over the defined
operations, in particular over SIMD instructions.  The first step is
to define a generic theory for words of size $k$, with the usual
arithmetic and bit-wise operations. The semantics of arithmetic
operations are based on two injections (signed and unsigned) into
integers and arithmetic modulo $2^k$. For bit-wise operations, we rely
on an injection to Boolean arrays of size $k$.  Naturally a link
between both representations (int and Boolean array) is also created,
which allows proving for example that shifting a word $n \ll i$ is the
same as multiplying it by $\textsf{to\_uint}~ 2^i$.

Scalar x86 operations are formalized using the theory for words, and
useful lemmas about the semantics of these instructions are also
proved as auxiliary lemmas. For example, the formalizations of
\textsf{shl} and \textsf{shr} permit proving lemmas like
$\mathsf{shl}~x~i \oplus \textsf{shr}~x~(k - i) = \textsf{rol}~x~i$,
under appropriate conditions on $i$.

The semantics of SIMD instructions rely on the theories for 128/256
bit words, but the semantics must be further refined to enable viewing
words as arrays of sub-words, which may be nested (e.g., instruction
\textsf{vpshufd} sees 256-bit words as two 128-bit words, each of them
viewed as an array of sub-words). To ease this kind of definition, we
have defined a bijection between words and arrays of (sub-)words of
various sizes.  Then vector instructions are defined in terms of
arrays of words.

\subsubsection*{Memory model}
\EasyCrypt does not provide the notion of pointer natively.  We rely
on the concept of a global variable in \EasyCrypt, which can be
modified by side effects of procedures, to emulate the global memory
of \Jasmin and the concept of pointer to this memory.  A dedicated
\EasyCrypt library defines abstract type \ec{global\_mem\_t} equipped
with two basic operations for load \ec{mem[p]} and store
\ec{mem[p \(\leftarrow\) x]} of one byte, as follows:

{\begin{lstlisting}[language=easycrypt,xleftmargin=0pt,xrightmargin=0pt,basicstyle=\footnotesize,frame=none]
type address = int.
type global_mem_t.
op "_.[_]" : global_mem_t -> address -> W8.t.
op "_.[_<-_]" : global_mem_t -> address -> W8.t -> global_mem_t.
axiom get_setE m x y w : m.[x <- w].[y] = if y = x then w else m.[y].
\end{lstlisting}}

From this basic axiom we build the semantics of load and store
instructions for various word sizes. The \Jasmin memory library then
defines a single global variable \ec{Glob.mem} of type
\ec{global\_mem\_t}, which is accessible to other \EasyCrypt modules and
is used to express pre-conditions and post-conditions on memory
states.

\subsubsection*{Soundness}
The embedding of a \Jasmin program into \EasyCrypt is sound, provided
the program is safe. This is because the axiomatic model of \Jasmin
in \EasyCrypt is intended to be verification-friendly, and assuming 
safety yields much simpler verification conditions and considerably 
alleviates verification of functional and equivalence properties. 
This assumption is perfectly fine, since \Jasmin programs are automatically 
checked for safety before being compiled and embedded into \EasyCrypt. 
As potential future work, it would be interesting to make our safety checker
certifying, in the sense that it automatically produces a proof of
equivalence between the \Coq and \EasyCrypt semantics of \Jasmin
programs---technically, this would be achieved by formalizing in \Coq
a simpler semantics for safe programs, and proving automatically that
the two semantics coincide for safe programs. The coincidence between
the simpler semantics in \Coq and the \Jasmin semantics would still
need to be argued informally.


\subsubsection*{Reusable \EasyCrypt libraries}
In the course of writing correctness proofs for our use cases we have
created a few \EasyCrypt libraries that will be useful for future
projects. In addition to the interchangeability of generic vectorization 
modules \ec{Ops} and \ec{OpsV} which we mentioned in Section~\ref{sec:motivex},
significant effort was put into enriching the theories of words in 
order to facilitate proofs of computations over
multi-precision representations.
Concretely, a theory was created that permits tight control over
the number of used bits within a word (a form of range analysis),
which is crucial for dealing with delayed carry operations and
establishing algebraic correctness via the absence of overflows.
The central part of this library is generic with respect to the number of
limbs, so that operations like addition and school-book multiplication
can be handled in a fully generic way (here we rely heavily on the
powerful ring theory in \EasyCrypt). When dealing with constructions
such as \poly, base on primes which are very close to a power
of $2$, this means that only the prime-specific modular reduction
algorithm needs special treatment.
Moreover, this theory was fine-tuned to interact well with SMT
provers, enabling the automatic discharge of otherwise tedious to
prove intermediate results.

\subsection{Verification of timing attack mitigations}
The \EasyCrypt embedding of \Jasmin programs is instrumented with
leakage traces that include all branching conditions plus all
accessed memory addresses (this also includes array indexes since
an access in a \textit{stack} array will generate a memory access at the
assembly level). It is then possible to check that the private inputs
do not interfere with this leakage trace in the classical sense that,
for all public-equivalent input states $x_1 \equiv_\mathsf{pub} x_2$,
the program will give rise to identical leakages $\ell_1 = \ell_2$.
Figure~\ref{fig:ctjasmin} shows an example of the generated
instrumented \EasyCrypt code.

\begin{figure}
\begin{jasmincode}
\jasminkw{fn} \jasmindname{store2}(\jasminstorageclass{reg} \jasmintype{u64} p, \jasminstorageclass{reg} \jasmintype{u64}[2] x) \{\\
\jasminindent{1}[p + 0] = x[0];\\
\jasminindent{1}[p + 8] = x[1];\\
\}
\end{jasmincode}
\input{listings/ctjasmin}
\caption{\label{fig:ctjasmin}%
  \EasyCrypt code (bottom) instrumented for {\em constant-time} verification
of a \Jasmin program (top).}
\end{figure}

Pleasingly, \EasyCrypt tactics developed to deal with information
flow-like properties handle the particular equivalence relation
associated with co-called {\em constant-time} security extremely effectively.
In particular, \EasyCrypt provides the \ec{sim} tactics which is
specialized on proving equivalence of programs sharing the same control
flow (which is the case here, as we are reasoning
about two executions of the same program). The tactic is based on
dependency analysis and also proved very useful in justifying simple
optimizations like spilling, which do not affect the control flow.
In the case of constant-time verification there is a very interesting
side-effect to the dependency analysis performed by this tactic:
it is able to infer sufficient conditions (equality of input variables) 
that guarantee equality of output variables. When applied to
constant-time verification this means that, when this tactic
is successful (which was the case for our use-cases) the user
just needs to check if the inferred set of variables are all
public.
We note that performing this kind of analysis at the assembly
level is usually hard. We take advantage of the fact that 
\Jasmin provides a high-level semantics that makes it suitable
for verification; in particular, the clear separation between 
memory, stack variables and stack arrays at source level greatly simplifies
the problem.

\section{Case Study: ChaCha20}
\label{sec:casestudies}

\noindent
We present the \Gimli
case study in Appendix~\ref{app:gimli}

\subsubsection*{Algorithm overview}
\chacha is a stream cipher, which we describe as specified 
in TLS 1.3. It defines an algorithm that expands a 256-bit key 
into $2^{96}$ key streams (each stream is associated with a 96-bit 
nonce) each consisting of $2^{32}$ blocks (each 64-byte block is 
associated with a counter value).%
\footnote{The typical composition with \poly, also adopted in TLS 1.3, uses
\chacha with counter $0$ to generate the key material for \poly;
the keystream generated for increasing counters starting at $1$
is used for encryption by {\sc xor}-ing with the plaintext.
\poly is then used to authenticate the ciphertext (prefixed with
any metadata that must also be authenticated) after adding a
length-encoding padding. We analyse the two algorithms
in isolation to facilitate comparison with other
implementations, and because the verification challenges are
significantly different.}
\chacha defines a procedure to transform an initial state into
a keystream block. The initial state is constructed using the 
256-bit key $k$ (seen as eight 32-bit words), 
the 96-bit nonce $n$ (seen as three 32-bit words), a 
32-bit counter $b$ and four 32-bit constants $c$.
Pictorially, the initial state can be seen as the following matrix, where
on the left-hand side we show the arrangement of 32-bit words
and on the right-hand side we show the matrix entry 
numbering.

\[
\begin{array}{ccc}
\begin{array}{cccc}
c & c & c & c\\
k & k & k & k\\
k & k & k & k\\
b & n & n & n
\end{array}
& \qquad &
\begin{array}{cccc}
0  & 1  & 2  & 3\\
4  & 5  & 6  & 7\\
8  & 9  & 10 & 11\\
12 & 13 & 14 & 15
\end{array}
\end{array}
\]

The state transformation, which is repeated for 10 rounds, is based
on the following operation that acts upon four 32-bit words at
a time:

\begin{center}
\begin{minipage}{.4\textwidth}
\underline{$\mathsf{Qround}(a,b,c,d)$:}\\
$
\begin{array}{l}
a \gets a + b; \\
c \gets c + d; \\
a \gets a + b; \\
c \gets c + d; \\
\end{array}
$ \  $
\begin{array}{l}
d \gets d \oplus a; \\
b \gets b \oplus c; \\
d \gets d \oplus a; \\
b \gets b \oplus c; \\
\end{array}
$ \ $
\begin{array}{l}
d \gets \textsf{rol}\ d\ 16; \\
b \gets \textsf{rol}\ b\ 12; \\
d \gets \textsf{rol}\ d\ 8;  \\
b \gets \textsf{rol}\ b\ 7;  \\
\end{array}
$ 

Return $(a,b,c,d)$
\end{minipage}
\end{center}

Each round updates the state by gradually modifying
the state, four words at a time using the \textsf{Qround}
function above, according to the following sequence of
4-word selections: 
$(0, 4, 8, 12)$,
$(1, 5, 9, 13)$,
$(2, 6, 10, 14)$,
$(3, 7, 11, 15)$,
$(0, 5, 10, 15)$,
$(1, 6, 11, 12)$,
$(2, 7, 8, 13)$ and
$(3, 4, 9, 14)$.
The final keystream block results from the {\sc xor}
combination of the output of the 10 rounds with the
initial state.

\subsubsection*{Our implementation}

We have defined and proved two versions of \chacha, one relying
only on scalar operations (no vectorization) and the second one relying
on {\sc avx2}.

The {\sc avx2} version combines two approaches to the 
optimization of \chacha: for short messages (up to 256 bytes) we 
follow the lines of~\cite{EPRINT:GolGue13}, whereas for large messages we adopt
the strategy of \textsf{OpenSSL}.
Both approaches were ported to \Jasmin, and further optimization
of instruction selection, scheduling and spilling was conducted
to obtain additional reductions in cycle counts.

Both approaches rely on vectorized instructions, but with different
parallelization approaches.
For small messages, two (for messages of up to 128-bytes) or 
four keystream blocks are computed at a time, as there are 
enough 256-bit registers available to enable the parallel 
computation of some steps within the same block using 
a dedicated state representation.\footnote{Four 256-bit registers 
are used to store two initial states for two successive counters, 
which permits computing four lines of code in \textsf{Qround} with 
only three vector instructions, simultaneously for the two states.
The round is completed by permuting the states, again using
vector instructions, and repeating the same technique to
compute the last four lines in \textsf{Qround}.}
For long messages, this no longer pays off due to the need 
for spills, and we rely on sixteen 256-bit registers,
which permit storing the states for 8 block computations
using a direct paralellisation approach that replicates
a fast implementation of a single block. 

In the next section we give detailed performance
benchmarks for our code, and compare to existing implementations.
Next, we describe how, in addition to being the fastest, our code
is also proved functionally correct. 

\subsubsection*{Formal verification}

The scalar and {\sc avx2} versions have (almost) the same
specification, which corresponds to the \HACL specification, with some
differences we present later.  Similarly to what we did for \poly, we
define an \EasyCrypt imperative reference implementation and show that
it satisfies \HACL functional specification using Hoare logic. Then,
we prove the equivalence between this reference implementation and
both of our optimized implementations.

The main challenge when proving correctness of the imperative
specification lies in memory operations. The imperative specification
stores ciphertext blocks eagerly (512-bits at a time),
while the functional specification stores the full ciphertext in one
go at the end. Therefore, we need a condition ensuring that stores do
not erase the fragment of the initial plaintext that remains to be
encrypted. 
Formally, we require that $ {\sf plain} + {\sf len}
\leq {\sf output} \vee {\sf output} \leq {\sf plain}$.

Proving equivalence with the scalar optimized implementation is relatively 
straightforward. 
The main difficulties come from optimizations of the memory operations. Indeed, in the optimized version we use 64-bit accesses whenever possible, instead of byte-level accesses as in the reference implementation. This allows to save spilling and to reduce the number of loads and stores by a factor of 8.

The proof of the {\sc avx2} version is more intricate.
There are two different implementations for short messages and long messages.
However, we adopt the same proof strategy in both cases. 
We describe the long message case.
First we change the control flow of the main loop, so that
each loop iteration computes 8 independent states.
Then, we lay the groundwork for vectorization:
rather than manipulating 8 arrays of sixteen 32-bit words, we now
manipulate sixteen arrays of eight 32-bit words (here we leverage
\EasyCrypt automation significantly). 
Finally, we prove the we can use {\sc avx2} instructions
to replace multiple scalar instructions.
Again, the main difficulty is to deal with optimized memory access
operations, which now uses 256-bit loads and stores.
At this point, the 8 states are represented by a $16\times 8$ matrix,
which needs to be transposed in order to be {\sc xor}-ed with the 
plaintext (using 256-bit operations) and stored in memory. 
For performance reasons, this is done in two steps, each dealing with
half of the matrix. Because of this, we need a slightly stronger restriction on the input and output pointers than in the scalar version. They need to be
either equal or to point to disjoint memory regions. Formally, we require that
(${\sf plain} = {\sf output} \vee 
   {\sf plain} + {\sf len} \leq {\sf output} \vee 
   {\sf output} + {\sf len} \leq {\sf plain}$).


\section{Benchmarks}
\label{sec:bench}

%
\subsubsection*{Methodology}
The performance evaluation of the \Jasmin implementations of \chacha and \poly was carried
out using the benchmarking infrastructure offered by {\sc supercop}, version 20190110. 
All measurements were performed on an Intel i7-6500U (Skylake) processor clocked at 2.5GHz, 
with Turbo Boost disabled, running Ubuntu 16.04, kernel release 4.15.0-46-generic.
The available compilers for all non-\Jasmin code were {\sc gcc} 8.1 and \CompCert 3.4. 
Unless explicitly stated otherwise, {\sc gcc} was used.

\subsubsection*{Baselines}
Our benchmarks compare the new \Jasmin implementations to the fastest implementations
for the same primitives and architecture in the following cryptographic libraries:
\textsf{OpenSSL}, \HACL and \textsf{Usuba}. 
We integrated external libraries in {\sc supercop} by compiling them into static
libraries and renaming symbols to remove naming collisions; this is particularly
important for libraries which we compiled using different compilers for 
comparison---for instance \HACL was compiled with both {\sc gcc} and
\CompCert.
A small patch to the {\sc supercop} benchmarking scripts was also added to
include these libraries in the set of evaluated implementations.
Finally, we created a binding to connect these implementations to the 
API that {\sc supercop} requires for evaluation. Concretely, we implemented
APIs \textsf{crypto\_stream\_xor} for \chacha and \textsf{crypto\_onetimeauth}
for \poly.

%
%

\subsubsection*{Results}
\begin{figure}[t]
\centering
      \includegraphics[width=.45\textwidth]{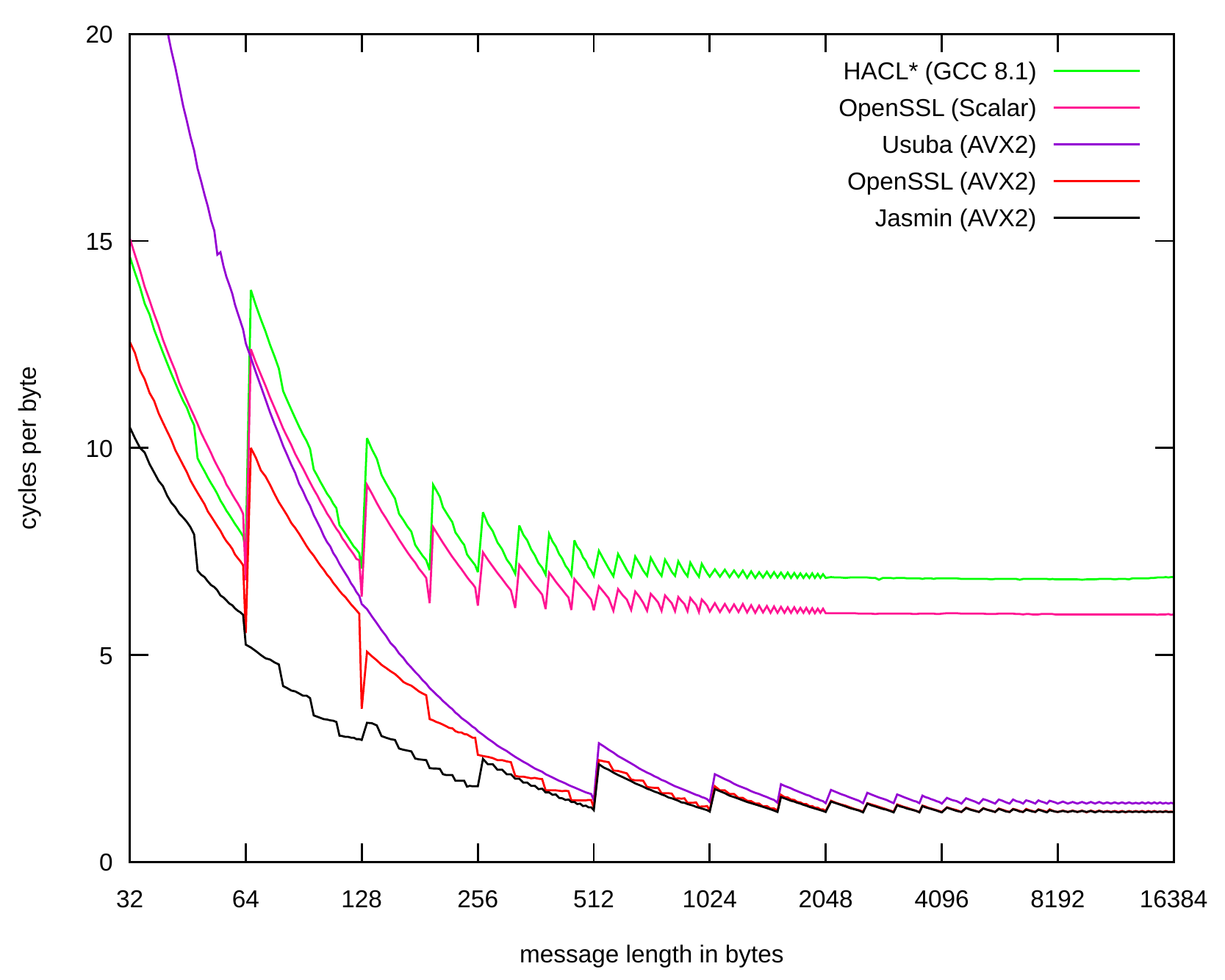}
      \includegraphics[width=.45\textwidth]{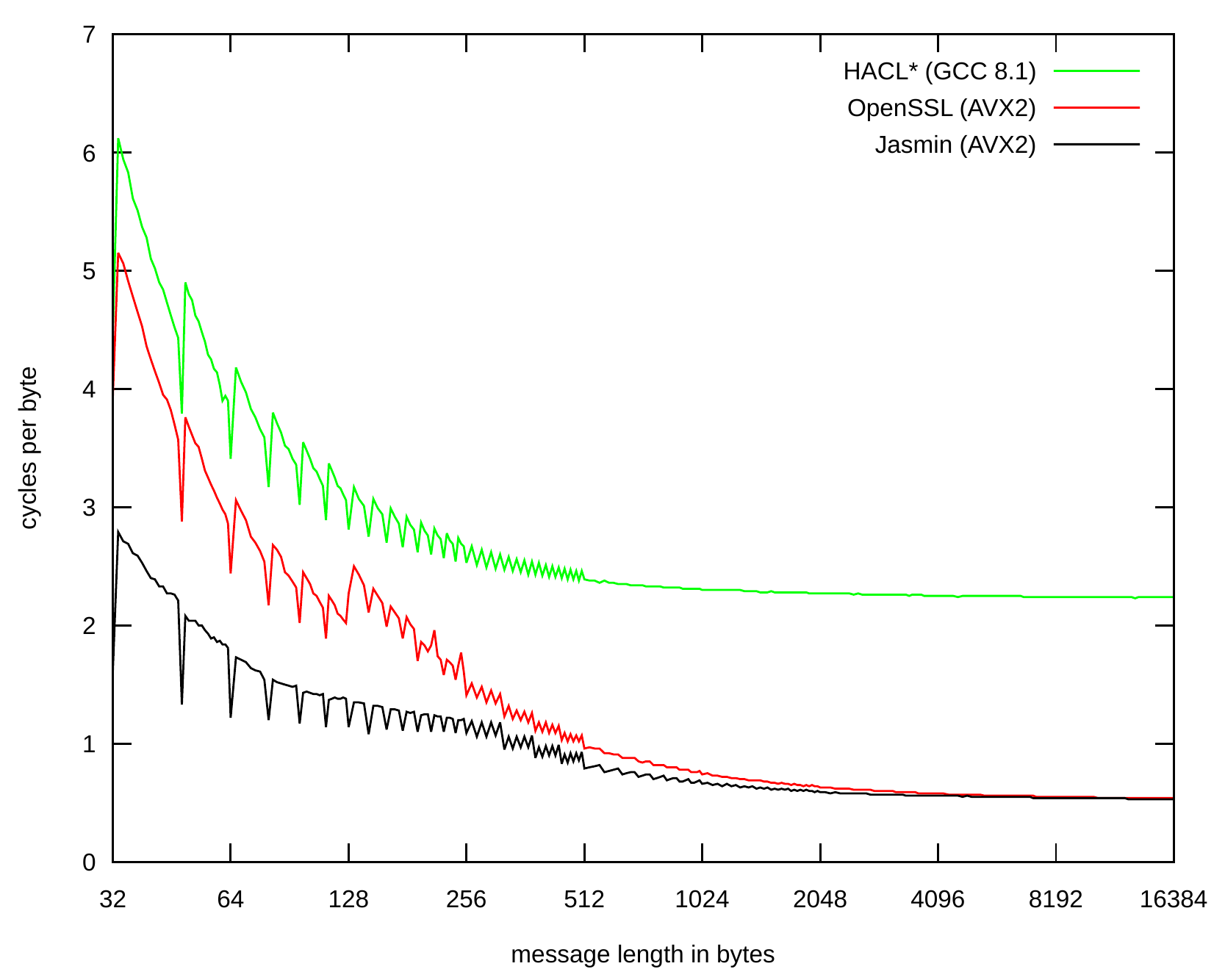}
      \caption{Comparison to non-verified code. \\ Top: \chacha. Bottom: \poly. }
      \label{fig:nonvcode}
      \vspace{-12pt}
 \end{figure}

Figure~\ref{fig:nonvcode} shows the
benchmarking results of our implementations of \chacha
and \poly in comparison to the prominent alternatives in
terms of performance. We emphasize that in this comparison our
code is the only one verified for functional correctness,
safety and so-called {\em constant-time} security (\HACL is compiled with 
non-verifed {\sc gcc}). 
The comparison with OpenSSL for small messages 
should be taken with a grain of salt, as there is some overhead 
due to binding with {\sc supercop} using the C API.

For \chacha the figure shows a clear difference between 
non-vectorized and vectorized code and our implementation essentially
matches OpenSSL as messages grow 
(we are measuring amortized cycles per byte).
In particular note that, for non-vectorized implementations,
the C code of \HACL is not much worse than OpenSSL's assembly.
The efficiency boost of vectorization is significant, even
for relatively small messages.
This gives relevance to our results, as we now
support fully verified vectorized assembly implementations. 

For \poly we compare to \HACL and OpenSSL's best implementation, which
has a structure similar to ours and uses 
non-vectorized code for small messages.
We can see that our implementation is again the fastest and, more
importantly, that vectorized code is once more crucial to make the most
of the computational platform (visible for large messages). 
Interestingly, the figure shows that OpenSSL seems to 
switch from non-vectorized to vectorized code at around 
128-byte messages, whereas our implementation does this at 256-bytes
and this seems to be advantageous.

\begin{figure}[t]
\centering
      \includegraphics[width=.45\textwidth]{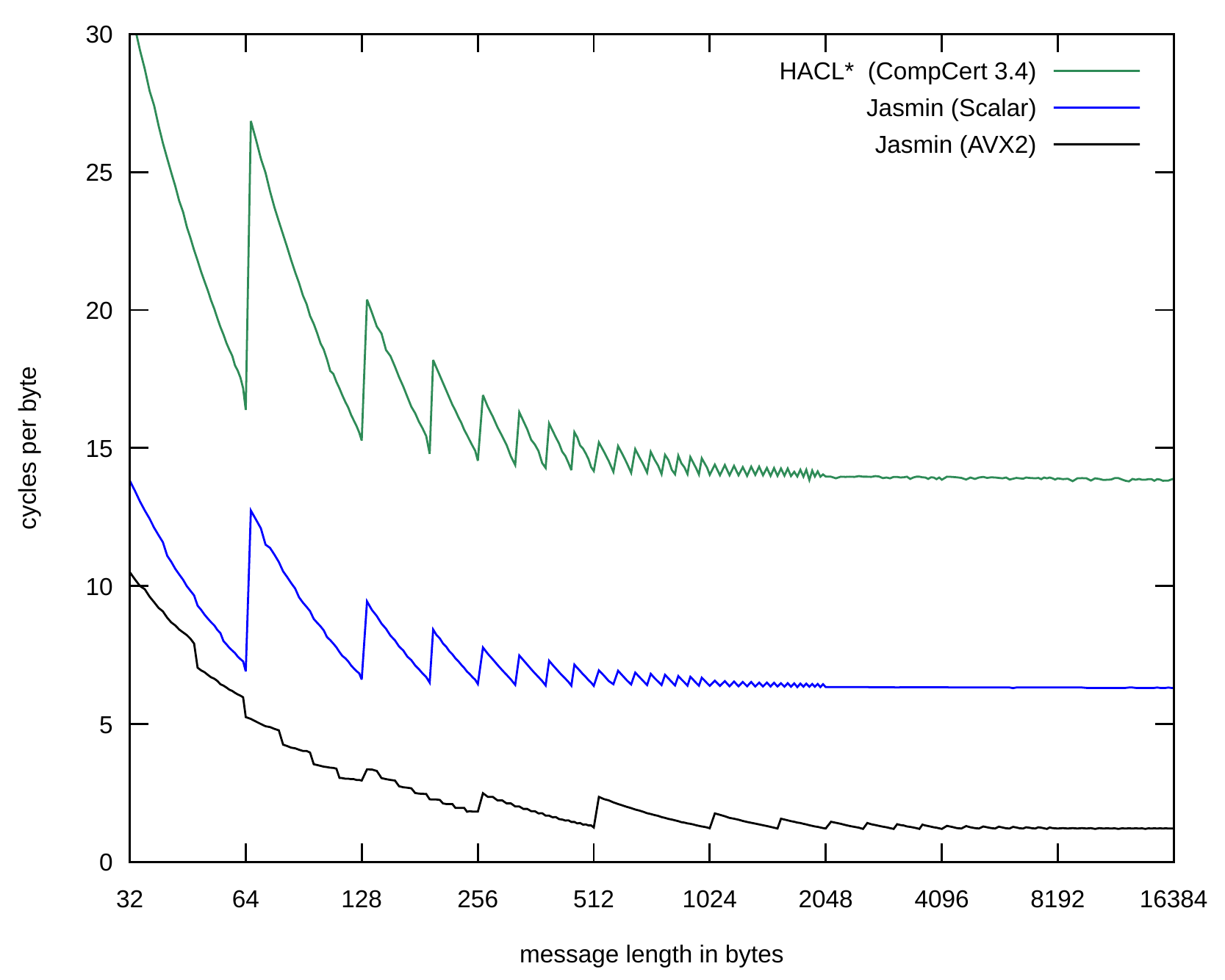}
      \includegraphics[width=.45\textwidth]{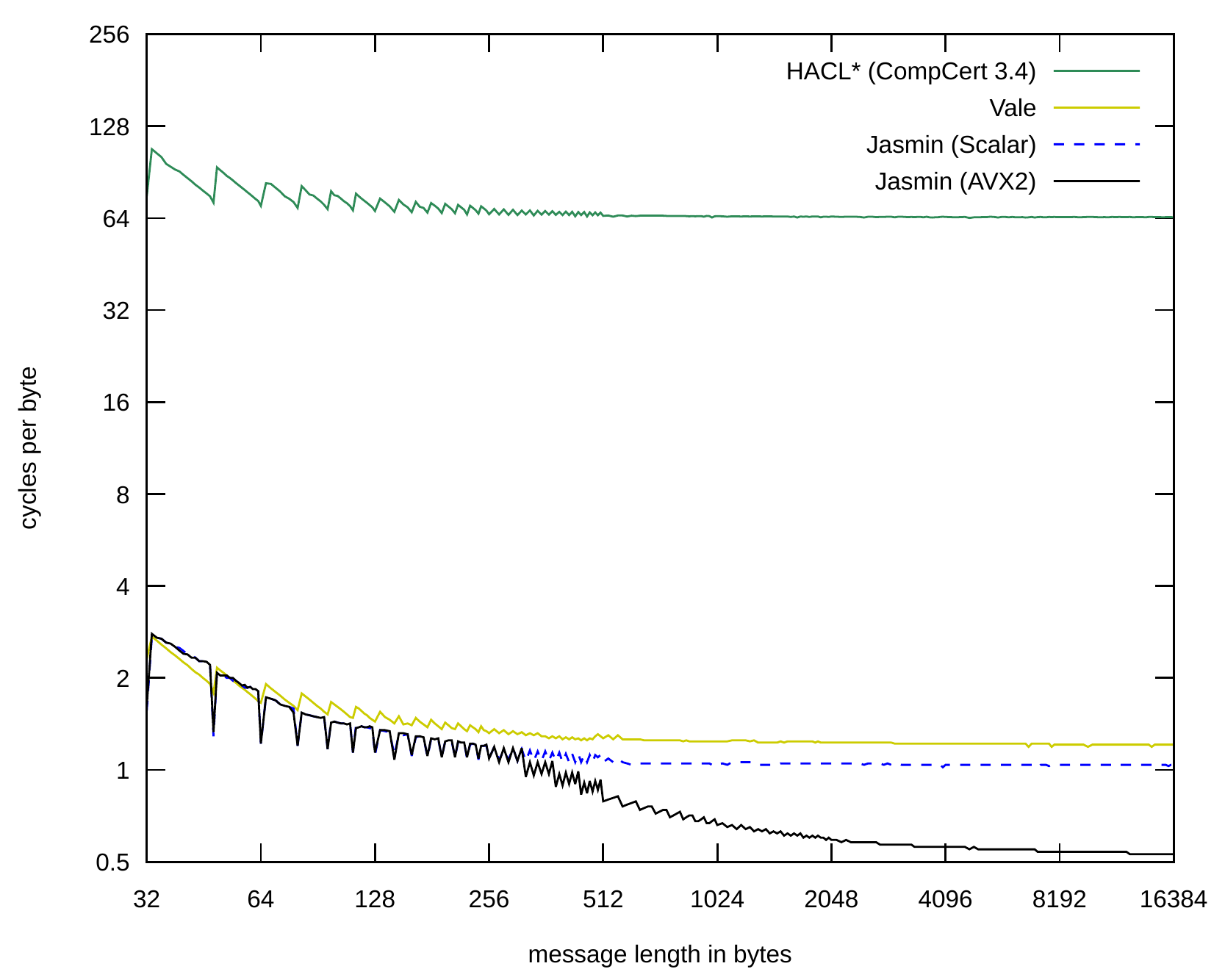}
      \caption{Comparison to verified code. \\ Top: \chacha. Bottom: \poly. }
      \label{fig:vcode}
      \vspace{-12pt}
 \end{figure}

Figure~\ref{fig:vcode} shows a
comparison to verified code, where \HACL is now compiled with
\CompCert. 
For \chacha, we show both our vectorized and
non-vectorized implementations, so as to demonstrate that
there is indeed a big advantage in bypassing the compiler, 
even if not relying on vectorization. Indeed, our non-vectorized
code is still roughly $\times 2$ faster than \HACL, while our vectorized 
code is about $\times 10$ faster.

For \poly we compare both to \HACL and
to non-vectorized OpenSSL code verified 
in the Vale framework~\cite{BondHKLLPRST17}
(here the comparison is assembly to assembly and so it is
precise).
The fine-tuning of our implementation shows in the comparison
to the Vale-verified OpenSSL code (the dashed line depicts 
non-vectorized Jasmin code even for large messages for comparison).
The difference to \HACL in this case is huge, both for non-vectorized
and vectorized code, and it is 
due to the intensive use of algebraic operations.

As a final note, we emphasize that we do not claim that
ours is the only verification framework that permits 
achieving such results: for example, the vectorized \poly code from 
OpenSSL from Figure~\ref{fig:nonvcode} could be verified 
using Vale or some other framework and closely match our
code's performance. 
The intended take away message from this section is rather 
that our methodology and framework permit achieving this for 
{\em new} implementations, which can incorporate ideas for 
speed optimization and functional correctness proofs from 
cryptographers and further fine-tune them using \Jasmin.

%


\section{Related Work}
\label{sec:related}

Appel and collaborators prove functional correctness of C
implementations of SHA-256~\cite{Appel15}, HMAC~\cite{BeringerPYA15}
and HMAC-DBRG~\cite{YeGSBPA17}. Their proofs are carried in the
Verified Software Toolchain~\cite{Appel14}, an interactive program
verification tool for C programs. Their verified implementations can
be compiled to assembly using CompCert~\cite{Leroy06}, a formally
verified optimizing compiler for C.  
Their work does not
analyze the penalty of using a verified compiler, nor does it provide
any guarantee with respect to side-channels.

Zinzindohoue \emph{et al.}~\cite{ZinzindohoueBPB17} develop HACL*, a
portable C library that implements many modern cryptographic
primitives. Implementations are written in \fstar~\cite{SwamyCFSBY11}, a
SMT-based verification-oriented language, and formally verified
against a readable mathematical specification. By enforcing that
secrets are used parametrically, it is also possible to guarantee that
\fstar programs are protected against side channels in the cryptographic
constant-time model~\cite{AlmeidaBBDE16}.  Verified \fstar implementations
are first compiled into C using Kremlin, a highly optimized compiler
from \fstar to C~\cite{ProtzenkoZRRWBD17}, and to assembly, using the
\clang compiler or the \CompCert compiler. Their library is deployed
in Mozilla Firefox, Wireguard and other popular products. 
Their evaluation shows their libraries to be as fast as unverified C libraries,
but for this one must assume that functional correctness and protection against
timing attacks is preserved by compilation.

In a similar way, Erbsen \emph{et al.}~\cite{Erbsen19} develop an
infrastructure for generating verified C implementations of elliptic
curve arithmetic from high-level descriptions written in the Coq proof
assistant. In addition, their generated code is protected against side
channel attacks in the program counter model~\cite{MolnarPSW05}, since
it does not contain any branching statement. As for HACL*, their
verified C implementations can be as efficient as the fastest
unverified C implementations, when compiled with a non-verified compiler.
FiatCrypto implementations are deployed in Chrome, Android and
other popular products.

Bond \emph{et al.}~\cite{BondHKLLPRST17} develop the Vale framework
for proving functional correctness and side-channel protection of
cryptographic primitives. Implementations are written in pseudo
assembly and annotated with logical annotations. For verification,
implementations are translated to the Dafny verifier and validity of
the annotations (which entails functional correctness) is checked
using SMT solvers. Verified implementations are then compiled from
Dafny to assembly. The performance of their code is similar to
\textsf{OpenSSL}.

Fromherz \emph{et al.}~\cite{FromherzGHPRS19} develop an approach
based on \fstar for proving correctness of C programs with inlined x64
assembly. Their approach is based on defining a deep embedding of x64
assembly and formally verifying an executable verification condition
generator for x64 assembly programs. The latter is used in combination
with \fstar verification condition generator for proving correctness
proofs of hybrid programs. The Kremlin compiler is then used to
generate C programs with inlined assembly, to get a performance 
similar to Vale.


Bo-Ying Yang \emph{et al}~\cite{CCS:CHLSTW14,CCS:TsaWanYan17} develop
highly automated tools for proving functional correctness of efficient
assembly implementations of elliptic curve cryptography. The strength
of their approach is a combination of proof assistants and automated
verification tools.

These approaches target functional correctness, as well
as side-channel resistance and provable security. In addition, many
works focus exclusively on side-channel resistance and/or provable
security. In particular, there exist several tools for proving
constant-time security~\cite{CCS:BBCLP14,RodriguesPA16,AlmeidaBBDE16}
or for making programs constant-time by
compilation~\cite{WuGSW18,fact19}. Finally, a recent work develops a
certified compiler~\cite{BartheGL18} that preserves constant-time
security. 



\section{Conclusion}
\label{sec:conclusion}
We have developed a practical framework to build high-assurance and
high-speed assembly implementations. We have shown
the benefits of our approach by manually optimizing and 
verifying functional correctness and security against timing attacks
of code for two primitives from the TLS 1.3. ciphersuite.

There are several important directions for future work. First, we
intend to verify a richer set of cryptographic primitives, including
all the primitives used in TLS 1.3. Second, we intend to develop a
translation validation approach for automating equivalence proofs
between reference and vectorized implementations. Third, we intend to
extend \Jasmin{} to support other architectures. In addition, it would
be of theoretical interest to develop a formally verified (relational)
verification condition generator in \Coq.

\bibliographystyle{plain}
\bibliography{dblp,abbrev3,crypto}

\appendices

\section{Overview of \Jasmin design choices}
\label{app:jasmin}

\Jasmin~\cite{CCS:ABBBGL17} is a language designed for building
efficient and formally verified cryptographic primitives within a
single language. This entails empowering \Jasmin programmers to use
different programming idioms for different parts of the
implementation, as shown in Figures~\ref{fig:polyopt},
\ref{fig:chacha20_shuffle} and~\ref{fig:bash_R4}.

On one hand, \Jasmin aims to provide the highest level of control and
expressiveness to programmers. Informally, the essential property that
\Jasmin aims to achieve is \emph{predictability}: the expert
programmer will be able to precisely anticipate and shape the
generated assembly code, so as to be able to achieve optimal
efficiency.  This means that the programmer must specify the storage
for program variables (stack, register) and must handle spilling
explicitly (the compiler will fail if it cannot find a spill-free
allocation).  \Jasmin also ensures that side-effects are explicit from
the program code by treating flags as boolean variables; this not only
gives explicit control over flags, but also makes verification of
functional correctness and even constant-time security significantly
simpler, as all non-memory-related instructions can be treated as pure
operators.

On the other hand, \Jasmin provides a uniform syntax that unifies
machine instructions provided by different micro-architectures. The
main purpose of this syntax is to ease programming and to enhance
portability.%
\footnote{Platform-specific instructions are also available and can be
  used whenever important, e.g., for efficiency. In particular,
  programmers may always use a \Jasmin dialect where there is a strict
  one-to-one mapping between \Jasmin instructions and assembly
  instructions.}  At the source level, stack variables and register
variables are interpreted simply as variables; the storage modifier is
only used as advice for register allocation. In particular, at this
level the memory is assumed to be disjoint from stack storage. The
compiler will later refine this model and conciliate the fact that
stack data must reside in memory as well.

\Jasmin supports (inlineable) function calls, which naturally leads to
a style of programming that favors modularity, and supports
high-level control-flow structures, rather than jumps.  \Jasmin also
supports functional arrays for describing collections of registers and
stack variables.  This notation leads to compact and intuitive code
and simplifies loop invariants and proofs of functional correctness.
Arrays are meant to be resolved at compile-time, and so they can only
be indexed by \emph{compile-time expressions}.
These can be used to describe statically unrollable \textsf{for}
  loops and conditional expressions, which permits replicating within
  the \Jasmin language coding techniques that are typically
  implemented using macros in C.

These choices have no impact on the efficiency of the generated code,
as low-level cryptographic routines usually have a simple control-flow,
which is easily captured by these high-level constructions.
Moreover, it considerably simplifies verification of functional
correctness, safety and side-channel security, and is critical to
leverage off-the-shelf verification frameworks, which are often focused
on high-level programs.

\section{Additional Case Study: Gimli}
\label{app:gimli}

\subsubsection*{Algorithm overview}

\Gimli~\cite{CHES:BKLMMN17} is a permutation designed to be used as a component
in the construction of block-ciphers, hash-functions, etc.
It operates on 384-bits, and is optimized to offer a good security/performance
trade-off across multiple platforms, including the deployment of countermeasures against
side-channel attacks.
It applies a sequence of 24 rounds to a 384-bit state, seen as a $3\times 4$
matrix of 32-bit words. Each round consists of three operations: 
\begin{enumerate}[leftmargin=*]
\item a non-linear layer implemented as a 96-bit fixed permutation, which is applied 
to each 3-word column and comprises bit-wise operations and entry swaps;

\item a linear mixing layer using two different matrix entry
permutations, one applied every fourth round and one every second
round;
\item a constant addition, applied every fourth round.
\end{enumerate}
What makes \Gimli an interesting example is that its specification is actually given
as imperative pseudocode, which we can write in \Jasmin at the same level of abstraction
as shown in Figure~\ref{fig:gimliref}.\footnote{We note that for \chacha and \poly the
original specifications are also given as pseudocode; however we chose to present
our reference specifications as being the ones used in \HACL for the sake of 
interchangeability. We believe the fact we can adopt both styles of specification
speaks for the versatility of our approach.}

\begin{figure}
\begin{jasmincode}\input{listings/gimliref.jazz}\end{jasmincode}
\caption{\Gimli reference implementation in \Jasmin.}
\label{fig:gimliref}
\end{figure}

\subsubsection*{Implementation and formal verification}
Our implementation of Gimli demonstrates the use of another set of instruction
extensions. As suggested in Gimli's proposal~\cite{CHES:BKLMMN17}, we rely on SSE, which provides 128-bit registers and allows for parallelization within a single block. In particular, we process the four columns in parallel in the non-linear part of each round. We chose this particular
parallelization approach because we are not optimizing \Gimli 
for a specific construction, but rather as a generic building block. 
Indeed, when \Gimli is used in specific constructions, 
parallelization across several blocks can be achieved using more powerful 
instruction extensions, supporting wider vectors.

The proof of the SSE version of \Gimli is comparatively simpler to our 
other examples.
In the vectorized version, the state is an array storing four 128-bit
values, each corresponding to a line in the matrix. 
The linear operations that permute entries within lines can 
be implemented using shuffle instructions {\sf vpshufd 0xB1} and {\sf vpshufd 0x4E}.
Proving the equivalence between the shuffle in 128-bits word and the reference
implementation is done by a simple reduction step, as \EasyCrypt's semantics
of x86 operations is computable.

A more intricate argument is needed to deal with the implementation of 
an equivalent of a {\sf rol} instruction for vectors, which does not
exist natively. This is based on a 24-bit rotation, which can be emulated
by permuting bytes using the {\sf vpshufb} instruction.
Proving the correctness of this requires switching the way 
we view 128-bits words between four 32-bits words and sixteen 8-bits words.
Again, the proof of this optimization is done by computation. 

\section{Additional support \Jasmin developers}


\Jasmin programs can be compiled to assembly, assembled and linked to other programs.
However, the \Jasmin compiler is partial: some valid programs may fail to be compiled.
For instance, for the sake of predictability, no temporary variables are introduced 
to compile expressions; also, register allocation does not introduce spilling and 
fails if not enough registers are available.
Moreover, the compiler correctness proof does not provide guarantees for unsafe
programs so, even if a program does compile, running the generated code is not a good
means to obtain feedback on the semantics of a source program.

To overcome these difficulties and be able to run \emph{specification 
programs} during development---\Jasmin programs that should be easy to read 
but may fail to be compilable or efficient---we have introduced an interpreter,
i.e., an executable small-steps semantics, of the \Jasmin language.

\end{document}